\documentclass{eptcs}
\providecommand{\event}{ACL2 2018} 
\usepackage{breakurl}             
\usepackage{underscore}           

\usepackage[nocompress]{cite}

\usepackage{amsmath}
\usepackage{amssymb}
\usepackage{amsbsy}
\usepackage{mathpartir}
\usepackage{tikz}
\usepackage{boxedminipage} 
\usepackage{booktabs} 

\newcommand{\imp}{\Rightarrow}

\newcommand{\ie}{\textit{i.e.}}
\newcommand{\etc}{\textit{etc}}
\newcommand{\eg}{\textit{e.g.}}
\newcommand{\konst}[1]{\ensuremath{\mbox{\sf{#1}}}}

\newcommand{\set}[1]{\{ {#1} \}}

\title{Using ACL2 in the Design of Efficient, Verifiable Data Structures for High-Assurance Systems\footnote{Approved for Public Release, Distribution Unlimited}}

\author{David Hardin\footnote{The views expressed are those of the authors 
and do not reflect the official policy or position of the Defense
Advanced Research Projects Agency (DARPA) or the U.S. Government.}
\institute{Advanced Technology Center\\Rockwell Collins\\
Cedar Rapids, IA USA}
\email{david.hardin@rockwellcollins.com}
\and
Konrad Slind
\institute{Advanced Technology Center\\Rockwell Collins\\
Bloomington, MN USA}
\email{konrad.slind@rockwellcollins.com}}

\def\titlerunning{Using ACL2 in the Design of Efficient, Verifiable
  Data Structures}
\def\authorrunning{D.S. Hardin \& K.L. Slind}

\begin{document}

\maketitle

\begin{abstract}
Verification of algorithms and data structures utilized in modern 
autonomous and semi-autonomous vehicles for land, sea, air, and space  
presents a significant challenge. Autonomy algorithms, \eg,  
route planning, pattern matching, and inference, are based on 
complex data structures such as directed graphs and algebraic data 
types.  Proof techniques for these data structures exist, but are 
oriented to unbounded, functional realizations, which are not
typically efficient in either space or time.  Autonomous systems 
designers, on the other hand, generally limit the space and time 
allocations for any given function, and require that algorithms
deliver results within a finite time, or suffer a watchdog timeout.  
Furthermore, high-assurance design rules frown on dynamic memory 
allocation, preferring simple array-based data structure 
implementations.  

In order to provide efficient implementations of high-level data
structures used in autonomous systems with the high assurance needed
for accreditation, we have developed a verifying compilation technique
that supports the ``natural'' functional proof style, but yet applies
to more efficient data structure implementations.  Our toolchain
features code generation to mainstream programming languages, as well
as GPU-based and hardware-based realizations.  We base the
Intermediate Verification Language for our toolchain upon higher-order
logic; however, we have used ACL2 to develop our efficient yet
verifiable data structure design.  ACL2 is  particularly well-suited
for this work, with its sophisticated libraries for reasoning about
aggregate data structures of arbitrary size, efficient execution of
formal specifications, as well as its support for 
``single-threaded objects'' --- functional datatypes
with imperative ``under the hood'' implementations.

In this paper, we detail our high-assurance data structure design 
approach, including examples in ACL2 of common algebraic data types 
implemented using this design approach, proofs of correctness for
those data types carried out in ACL2, as well as sample ACL2
implementations of relevant algorithms utilizing these efficient,
high-assurance data structures.

\end{abstract}

\section{Introduction}

As autonomous systems have matured from laboratory curiosities to 
sophisticated platforms poised to share our roadways, sea lanes, and 
airspace, accrediting agencies are faced with the significant
challenge of verifying and validating these systems to ensure that 
they do not constitute a significant societal risk. Leaving aside the 
issues with the verification and validation of deep learning, even 
basic autonomy algorithms, \eg, route planning, pattern matching, 
and inference, are based on complex data structures, such as directed 
graphs and algebraic data types.  Proof techniques for these data 
structures exist, but are oriented to unbounded, functional
realizations, which are not typically efficient in either space or time.

Autonomous systems designers, on the other hand, generally limit the
space and time allocations for any given function, and require that
algorithms deliver results within a finite time, or suffer a watchdog
timeout.  Furthermore, high-assurance design rules, such as mandated
by RTCA DO-178C Level A \cite{DO-178C} for flight-critical systems, 
frown on dynamic memory allocation, preferring simple array-based 
data structure implementations.

In order to provide efficient implementations of high-level data
structures used in autonomous and other critical systems with the 
high assurance needed for accreditation, we have been developing a 
verification technique that supports the ``natural'' functional proof
style, but yet applies to more efficient data structure
implementations.  We have used ACL2 to prototype and refine an 
in-place data structure representation amenable for formal proof, 
as detailed in Section~\ref{RASLDASL}.  Applications include 
path planning on high-level graphs, inference engines for 
autonomous mission executives, \etc.

We have significant experience in the formal verification of practical
engineering artifacts \cite{Hardin2010}.  Our particular 
experience with the formal verification of array-based
implementation of algebraic data types indicates that these proofs are
extremely difficult, whether one takes a theorem proving
\cite{ACL2GPU}, model checking \cite{Hardin2009b}, or symbolic execution
\cite{Hatcliff2011} approach.  The former approach, while
yielding a proof for arbitrary array size, easily gets bogged down in
supplementary details (even when using list/array "bridging"
constructs such as ACL2 single-threaded objects \cite{stobj}), whereas
the latter two approaches suffer from lack of scalability, leading to
timeouts except for very small arrays --- but unsurprisingly,
most practical algorithms operate routinely on thousands
of data elements.  Therefore, we need a different approach.

We have applied verifying compiler technology to this problem.
Algorithms are expressed in a Domain-Aware System Language.  
A Domain-Aware language is one whose features are informed by
the requirements of the domain, such as primitive operations, 
environmental constraints, assurance and accreditation requirements, 
and so on, but which is not overtly tied to a particular 
domain.  Our primary goal is to craft a verification-enhanced 
programming language to support the high assurance requirements 
of the domains in which it will be used.

\section{Verification-Enhanced Programming Languages}

A \emph{verification-enhanced programming language} is one in which
properties of programs in the language can be formally stated and
reasoned about in an integrated environment. Examples of such
languages are SPARK/Ada \cite{sparkapps}, Dafny \cite{dafny:icse}, Guardol
\cite{guardol:tacas}, and certain C dialects\footnote{Note that any logic capable of
  expressing computable functions, \eg, ACL2 \cite{acl2:book} or
  higher order logic, can be regarded as a verification-enhanced
  programming language; our emphasis here is on more conventional
  programming languages.} \cite{leroy:cacm,typesbytes}.  Reasoning 
environments for C have become highly developed because of the 
pervasive usage of C in important system infrastructure such as 
operating systems, cryptography libraries, \etc.; but C does not 
support our goal of producing high-level functional correctness proofs.

\begin{figure}
\center{
\begin{tikzpicture}[auto]
\node (IDE) at (0,1.5) [shape=rectangle,draw]{IDE};
\node (AST) at (2,1.5) [shape=rectangle,draw]{AST};
\node (IVL) at (4,1.5) [shape=rectangle,draw]{IVL};
\node (SOLVERS) at (6.5,1.5) [shape=rectangle,draw]{SOLVERS};
\node (CODE) at (2,0) [shape=rectangle,draw]{CODE};
\path [->,thick] (IDE) edge  [loop above] node{\textit{parse; edit}} ();
\path [->,thick] (IDE) edge  [loop below] node{\textit{typecheck}} ();
\path [->,thick] (AST) edge [loop above] ();
\path [->,thick] (IVL) edge [loop above] ();
\draw [->,thick] (IDE) to (AST);
\draw [->,thick] (AST) to (IVL);
\draw [->,thick] (AST) to (CODE);
\draw [->,thick] (IVL) to (SOLVERS);
\draw [->,thick] (SOLVERS) to [out=135,in=45] node[swap]{\textit{counterexamples}} (IDE);
\end{tikzpicture}}
\caption{Verification tool design pattern.}
\label{vep:diagram}
\end{figure}
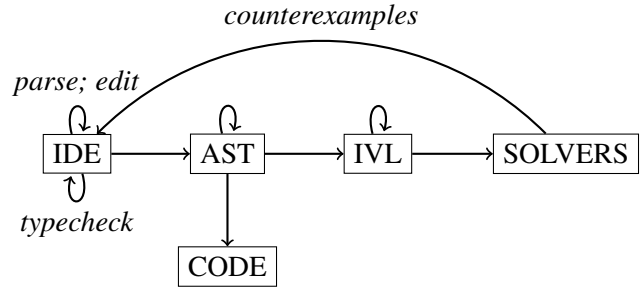

Once one has constructed a verification-enhanced language, it is
relatively simple to create another: after a point, much of the middle
and backend processing changes little, leaving only the frontend to be
adjusted to the new concrete syntax and type system. This explains the
rise of the \emph{intermediate verification language}. For example,
Boogie \cite{Boogie-IVL} underlies a number of languages, such as VCC
and Dafny, and WhyML \cite{why3} underlies verification-enhanced
versions of C and Ada.  Figure~\ref{vep:diagram} illustrates the basic
toolchain pattern: an IDE is used to create and edit programs and
their specifications; typechecking and semantic checking of programs
usually takes place here as well, in order to give good feedback on
program construction. The program is then mapped to an abstract syntax
tree (AST) representation suitable for code generation/compilation or
for mapping into a verification-friendly format. Source-to-source
transformations are possible in the AST representation and also in the
IVL representation.\footnote{In some systems, the AST representation
  and the IVL representation coincide.} From the IVL
representation, SMT solvers, or other automatic methods can be
invoked. Interactive proofs can also be initiated for properties
lying outside the domain of automation. In case the backend solvers
provide counterexamples, they may be translated into user-friendly
format and passed back to the IDE.

We note the following benefits of providing the IVL with a formal
operational semantics:

\begin{enumerate}


\item The semantics provide a formal basis for individual programs to
  be proved correct. By use of \emph{decompilation into logic}
  \cite{myreen:phd}, imperative programs in the operational semantics
  can be converted to equivalent logic functions by deduction. 

\item The formal semantics are the basis for proving the correctness of
  AST-to-AST transformations.  

\item An exciting aspect of the CakeML verified compiler
  \cite{cakeML:popl14} is the existence of a verifying translation
  from entities in the HOL logic to CakeML programs
  \cite{so-called-translator}.  As a consequence, HOL functions can be
  defined and have properties proved about them, then can be
  automatically translated to CakeML.  The correctness of the CakeML 
  compiler ensures that the behavior of the compiled binary version of
  the function is that of the original logic function. We can further 
  leverage this capability to establish a formal connection between 
  the AST for the original program and the final executable.

\end{enumerate}

\section{Modelling Imperative Languages}

\begin{figure}
\center{
\begin{tikzpicture}[auto]
\node (IDE) at (0,2.5) [shape=rectangle,draw]{IDE};
\node (HOL) at (4,2.5) [shape=rectangle,draw]{AST/IVL (HOL)};
\node (RADA) at (8,2.5) [shape=rectangle,draw]{RADA};
\node (Ada) at (0.25,0) [shape=rectangle,draw]{Ada};
\node (Java) at (1.75,0) [shape=rectangle,draw]{Java};
\node (CakeML) at (3.5,0) [shape=rectangle,draw]{CakeML};
\node (CUDA) at (5.5,0) [shape=rectangle,draw]{CUDA};
\node (VHDL) at (7.25,0) [shape=rectangle,draw]{VHDL};
\path [->,thick] (IDE) edge  [loop above] node{\textit{parse; edit}} ();
\path [->,thick] (IDE) edge  [loop below] node{\textit{typecheck}} ();
\draw [->,thick] (IDE) to node{\textit{formalize}} node [swap] {\textit{program}} 
                 (HOL);
\path [->,thick] (HOL) edge  [loop above] node{\textit{transform}} ();
\draw [->,thick] (HOL) to (Ada);
\draw [->,thick] (HOL) to (Java);
\draw [->,thick] (HOL) to (CakeML);
\draw [->,thick] (HOL) to (CUDA);
\draw [->,thick] (HOL) to (VHDL);
\draw [->,thick] (HOL) to node{\textit{automate}} node [swap] {\textit{proof}} 
                 (RADA);
\end{tikzpicture}}
\caption{DASL toolchain.}
\label{DASL:toolchain}\end{figure}
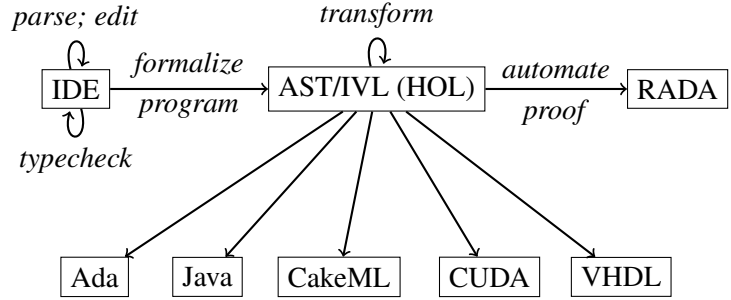

We will use \emph{DASL} as an example of a verification-enhanced
language.  DASL (Domain-Aware System Language) is a first order
hybrid functional/imperative language with constructs familiar from 
Ada, ML, Swift and other such languages. The
constructible types build on a standard collection of base types:
booleans, signed and unsigned integers (both bounded and unbounded), 
characters, and strings. Arrays and records support aggregation and
ML-style recursive datatypes provide tree-shaped data. Programs are
built from assignment, sequencing, conditional
statements, and (possibly recursive) procedures. The main novelty in
the statement language is support for ML-style \verb+match+
statements over the construction of datatypes. The concrete syntax is
conventional and should not cause any surprises. The instantiation of
the verification tool architecture of Figure~\ref{vep:diagram} to DASL
is shown in Figure~\ref{DASL:toolchain}.

We chose HOL4 as the environment for the ``middle-end'' of the 
DASL compilation toolchain due to its support for higher-order 
logic, the primary DASL developer's HOL4 expertise, as well as 
the verified path from executable logic function to binary code 
provided by the HOL4-based CakeML toolchain.

Once in the AST format, source code can be generated for a variety of 
programming languages. The backend solver is RADA, a SMT-based system 
for reasoning about recursive programs over algebraic datatypes
\cite{hung:phd}.

\subsection{Formal Operational Semantics}\label{opsem}

The operational semantics of DASL describes program evaluation by a
`big-step' inductively defined judgement saying how statements alter
the program state. The formula $
\mathsf{STEPS}\;\Gamma\;\mathit{code}\; s_1\; s_2$ says ``evaluation
of statement $\mathit{code}$ in environment $\Gamma$ beginning in
state $s_1$ terminates and results in state $s_2$''. We have also 
formalized a small-step semantics and proved
equivalence of the two semantics. Note that $\Gamma$ is an environment
binding procedure names to procedure bodies. We follow an approach
taken by Norbert Schirmer \cite{Schirmer-PhD}, wherein he constructed
a \emph{generic} semantics capturing a large class of sequential
imperative programs, and then instantiated the generic semantics to a
given programming language.

\subsection{Decompilation into Logic}

The pioneering work of Myreen \cite{myreen:phd} introduced the idea of
decompiling assembly programs to higher order logic functions; we have
adapted his approach to our high-level imperative language. For us, a
decompilation theorem has the stylized form

\[
\begin{array}{l}
\vdash \forall s_1\; s_2.\ \forall x_1 \ldots x_k. \\
\qquad  s_1.\mathit{proc}.v_1 = x_1 \land \cdots \land s_1.\mathit{proc}.v_k = x_k \ \land \\
\qquad  \mathsf{STEPS}\ \Gamma\ \boxed{\mathit{code}}\ (\konst{Normal}\; s_1)\ (\konst{Normal}\; s_2) \\
\qquad   \imp \\
\qquad   \mathtt{let}\; (o_1,...,o_n) = \boxed{\mathit{f}(x_1,\ldots,x_k)} \\
\qquad   \mathtt{in}\; s_2 = s_1\ 
           \mathtt{with} \set{\mathit{proc}.w_1 := \mathit{o}_1, \ldots, 
                              \mathit{proc}.w_n := \mathit{o}_n} \\
\end{array}
\]

\noindent
which essentially states that evaluation of $\mathit{code}$ implements
HOL function $\mathit{f}$. The antecedent $s_1.\mathit{proc}.v_1 =
x_1 \land \cdots \land s_1.\mathit{proc}.v_k = x_k$ binds logic
variables $x_1 \ldots x_k$ to the values of program variables $v_1
\ldots v_k$ in state $s_1$.  These values form the input for the
so-called \emph{footprint} function $\mathit{f}$, which delivers the
output values $o_1,...,o_n$ that are used to update $s_1$ to
$s_2$.\footnote{In our modelling, a program state is represented by a
  record containing all variables in the program. The notation
  $\mathit{s.proc.v}$ denotes the value of program variable $v$ in
  procedure $\mathit{proc}$ in state $s$.  The \texttt{with}-notation
  represents record update.}  One can see that a decompilation theorem
is a particular kind of Hoare triple. (An explicit Hoare triple
approach is used by Myreen in his work.)

\textbf{NB}. The footprint function $f$ is automatically synthesized
from $\mathit{code}$ and the decompilation theorem is proved
automatically. In other words, \emph{decompilation is an algorithm}:
it always succeeds, provided that all footprint functions arising from
the source program terminate.

Decompilation can also be applied to the problem of creating goals
from program specifications. A DASL specification sets
up a computational context---a state---and asserts that a property
holds in that state. In its simplest form, a specification looks like

\[
\begin{array}{l}
\mathtt{spec}\ \mathit{name}\ \{ \\
\quad \mathtt{var}\ <\!\mathit{local\ variable\ declarations}\!> \\
\quad \mathtt{in} \ \ \mathit{code};  \\
\quad \phantom{\mathtt{in}} \ \ \mathtt{check}\ \mathit{property}; \}\\
\end{array}
\]

\noindent
where $\mathit{property}$ is a boolean expression. A specification
declaration is processed as follows. First, suppose that execution of
$\mathit{code}$ starts normally in $s_1$ and ends normally in $s_2$,
\ie, assume 

\[
\begin{array}{l}
\mathsf{STEPS}\ \Gamma\ \mathit{code}\ (\mathsf{Normal}\;
s_1)\ (\mathsf{Normal}\; s_2). 
\end{array}
\]

We want to show that
$\mathit{property}$ holds in state $s_2$. We decompile $\mathit{code}$
to footprint function $f$ and decompile 
$\mathit{property}$ to footprint function $g$; then, formally, we need to show 

\[
\begin{array}{l}
(\mathtt{let}\; (\mathit{o}_1,...,\mathit{o}_n) = \mathit{f}(x_1,\ldots,x_k) \\ 
\mathtt{in}\; s_2 = s_1\ \mathtt{with}\ \set{\mathit{name}.w_1 := \mathit{o}_1, \ldots, \mathit{name}.w_n := \mathit{o}_n}) \\
\imp 
g\ s_2
\end{array}
\]

\noindent
The proof proceeds using facts about $f$, principally
its recursion equations and induction theorem, to show the translated
property $g$ holds on values projected from the final state $s_2$. The
original code and property have been freed---by sound deductive
steps---from the program state and operational semantics.

\section{Data Structure Compilation}

Compilation takes a DASL package expressed as a collection of type and
procedure declarations, and maps it, when a fixed size declaration is
present, to another package where algebraic
datatypes have been replaced by array-based representations, and
functions over the datatypes have been similarly lowered.



The compilation of data structures is phrased as a
source-to-source translation on DASL ASTs. It bears some resemblance
to the compiler described in \cite{tolmach:ML}. The target
representation was originally used in graph algorithms on GPUs, but
also serves quite well for imperative algebraic datatypes. 

\subsection{Target Representation}

Our array-based data structure representation is adapted from work 
by Harish and Narayanan \cite{Harish2007} on efficient graph 
algorithms for GPUs using the CUDA language; this layout was 
previously ported to ACL2 single-threaded objects \cite{ACL2GPU}. 
This design supports graphs with data at vertices and also on
edges. Graphs are bounded: both the number of vertices and the number
of outgoing edges per vertex are bounded.

One modification we have made to the layout described in the 
Harish and Narayanan paper \cite{Harish2007} is that zero designates 
a null vertex index or null edge index.  We explicitly allocate a zeroth element
for each array, and prove that the zeroth elements are unchanged 
by any array mutator.

A graph with maximum number of vertices $N$ and maximum number of outgoing edges
per vertex $M$ is represented by a seven-tuple 
$\mathit{store} = (V,D,E,W,\mathit{Vhd},\mathit{Vtl},\mathit{Vcount})$ where
\begin{itemize}
\item $V$ is the \emph{vertex} array of length $N$+1;
\item $D$ is the \emph{data} associated with each vertex, requiring an array of length $N$+1;
\item $E$ is the \emph{edge} array of length $MN$+1;
\item $W$ holds the ``weight'' or ``label'' data associated with each edge, also an array of length $MN$+1;
\item $\mathit{Vhd}$ is the index of the ``head'' vertex in $V$;
\item $\mathit{Vtl}$ is the index of the ``tail'' vertex in $V$; and 
\item $\mathit{Vcount}$ is the number of non-zero
  elements in $V$.
\end{itemize}

\begin{figure}
\begin{center}
\includegraphics[scale=0.5]{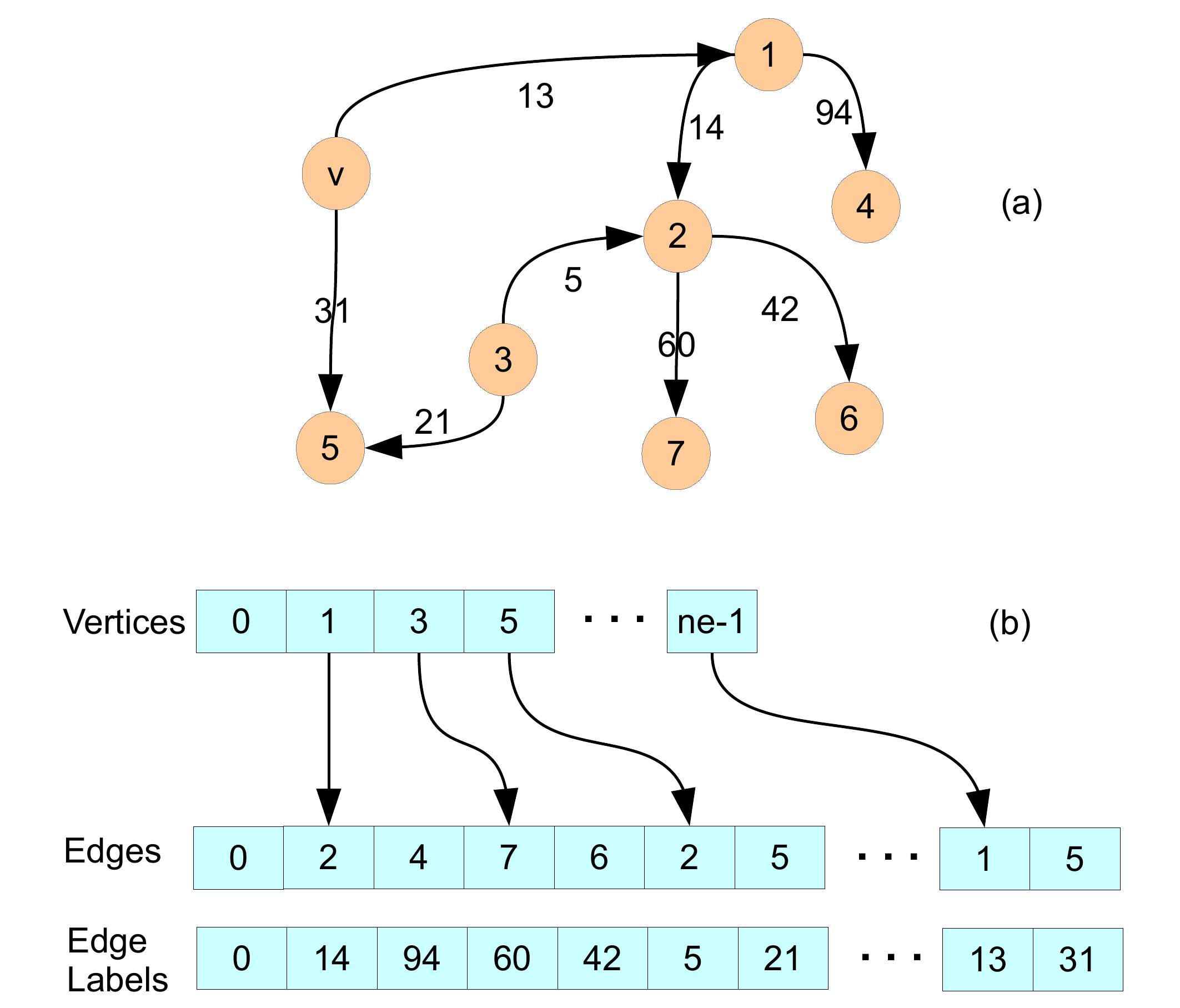}
\end{center}
\caption{(a) Graph fragment with two edges per vertex; (b) Array-based layout for the graph
  fragment.}
\label{graph-array-layout}
\end{figure}

\noindent The vertex array contains indices into the edge array, whereas the
edge array contains vertex indices, as shown in 
Fig.~\ref{graph-array-layout} (the data associated with each vertex
are not shown, in the interest of clarity).  The weight array contains 
the weight of each edge, and thus is the same size as the edge array.  
Note that this basic graph tuple can be expanded to include additional 
data, \eg, keys for keyed data structures.

\subsection{Verification}

To justify the translation into the array-based representation is an
exercise in compiler verification. Our approach uses decompilation
into logic. We also want to join the compiler verification results
with properties of individual programs. For example, programs over
recursive types in embedded systems are typically tail
recursive, in order to avoid memory allocation. However, tail
recursive programs (and their notorious \texttt{while}-loop brethren) are 
harder to reason about than recursive versions. So it is desirable to
also have a high-level logical characterization of the program to
reason about, provided properties proved at that level can be
transported to the program representations. 

\begin{figure}
\center{
\begin{tikzpicture}[auto]
\node (IMP) at (2,2) [shape=rectangle,draw]{$P_{\mathit{imp}}$};
\node (ARRAY) at (6,2) [shape=rectangle,draw]{$P_{\mathit{array}}$};
\node (IMP-FN) at (2,0) [shape=rectangle,draw]{$\mathit{Fn}_{\mathit{imp}}$};
\node (ARRAY-FN) at (6,0) [shape=rectangle,draw]{$\mathit{Fn}_{\mathit{array}}$};
\node (LOGIC-FN) at (0,0) [shape=rectangle,draw]{$\mathit{Fn}_{\mathit{logic}}$};
\node (CAKEML) at (4,-2) [shape=rectangle,draw]{CakeML};
\draw [->,thick] (IMP) to node {\textit{compilation}} (ARRAY);
\draw [->,thick] (IMP) to node [swap]{\textit{decompilation}} (IMP-FN);
\draw [->,thick] (ARRAY) to node [swap] {\textit{decompilation}} (ARRAY-FN);
\draw [-,thick] (LOGIC-FN) to node [swap] {$\equiv$} (IMP-FN);
\draw [-,thick] (IMP-FN) to node [swap] {$\equiv$} (ARRAY-FN);
\draw [->,thick,dashed] (IMP-FN) to (CAKEML);
\draw [->,thick,dashed] (ARRAY-FN) to (CAKEML);
\end{tikzpicture}}
\caption{Verifying translation.}
\label{translation-verification}
\end{figure}
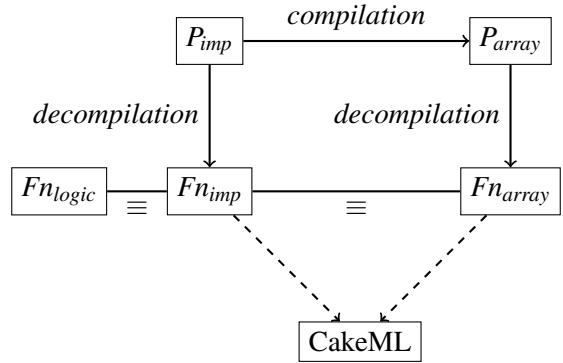

Figure~\ref{translation-verification} illustrates the relationships:
imperative program $P_\mathit{imp}$ over tree-structured data is 
compiled to program $P_\mathit{array}$ over our array-based 
representation.  These are decompiled to $\mathit{Fn}_{\mathit{imp}}$ 
and $\mathit{Fn}_{\mathit{array}}$, respectively.  In our design, the 
formal equivalence between $\mathit{Fn}_{\mathit{imp}}$ and 
$\mathit{Fn}_{\mathit{array}}$ is intended to be automatically 
proved. The tactics for this are in development.  On the other hand,
the relationship between high-level logical function
$\mathit{Fn}_{\mathit{logic}}$ and $\mathit{Fn}_{\mathit{imp}}$ may
require ingenuity to show. The end result is that properties
proved of $\mathit{Fn}_{\mathit{logic}}$ can be transported---by use of the
equivalence theorems---so that they apply, modulo adjustments to the
underlying representation, to the evaluation of $P_\mathit{array}$ on
the array representation.


Another interesting point is that the translation from
$P_\mathit{imp}$ to $P_\mathit{array}$ is \emph{informal}, \ie,
compilation is not achieved by proof steps, and yet decompilation into
logic still allows a formal relationship between $P_\mathit{imp}$ and
$P_\mathit{array}$ to be shown.

\section{Graphs}

The support for imperative functional programming via compilation into
the array-based representation offers a useful level of abstraction
for many programs. However, there are important applications where the
full expressiveness of graphs is needed, for example route
planning. In such cases, graphs could certainly be encoded via
datatypes, but it is much more appealing to provide a direct
representation.  Therefore we have added graph constructs as 
primitives in DASL.

In particular, the declaration

\[ \konst{graphtype}\; \mathit{name} \; (\konst{nodeLabel}, \konst{edgeLabel}) \]

\noindent 
creates a new type $\mathit{name}$ with the specified types of node
labels and edge labels. With that declaration, an API for creating,
traversing, and updating graphs of type $\mathit{name}$ becomes
available to the programmer. To exactly specify bounds on the
numbers of nodes and edges, the {\small\verb+sized+} declaration is
extended, thus

\[ \konst{sized}\ \mathtt{theGraph} : \mathit{name}\; (m,n) \]

\noindent 
declares a global variable {\small\verb+theGraph+} holding an empty
graph of type $\mathit{name}$ with at most $m$ nodes, where each node
can have at most $n$ outgoing edges.

As an example of the use of DASL to implement the sorts of graph
algorithms commonly encountered in autonomy and other 
high-assurance applications, consider the code for 
depth-first search depicted in Figure~\ref{fig:dfs}. The type of 
graphs handled by this code has unsigned integers at nodes (vertices) and 
unsigned integers labelling edges; other vertex and edge label types 
could just as easily been declared.  Constants used to bound the graph 
are also declared. Finally, a fine point of the design is whether to
expose the type of vertices.  Since vertices are used to index into
various data structures, we do show the representation.

{\footnotesize
\begin{verbatim}
  type vertexLabelTy = uint;
  type edgeLabelTy = uint;

  graphtype graph (nodeLabel = vertexLabelTy, 
                   edgeLabel = edgeLabelTy);
  const MAX_NODES : uint = 50000;
  const MAX_EDGES : uint = 4;
  type vertex = uint;

  sized theGraph : graph (MAX_NODES, MAX_EDGES);
\end{verbatim}}

\noindent 
In this variant of depth-first search, we capture the spanning
tree of the depth-first search as we proceed from a starting vertex.
We use a binary search tree (BST) of key/value pairs to capture the 
spanning tree as the depth-first search proceeds, where the keys 
are the reached vertices, and the values are the immediately 
preceding vertices for the reached vertices.  The BST for the 
spanning tree serves a dual purpose, as it is also used to check 
(via the \emph{exists} function) whether a given vertex has been 
previously encountered during the current search, so that we can 
cut off search to already-explored sub-graphs.

The search function {\small\verb+DFS_span+} is a recursive function
whose parameters include a target vertex to be searched for
(\emph{target}); the graph \emph{G}; the emerging \emph{spanning_tree}  
of type BST; and a current \emph{fringe} of unexplored edges,
represented as a list of \emph{(vertex, predecessor-vertex)} pairs 
(this list is conceptually a stack, and could be implemented using a defined DASL
stack datatype).  We also implement a ``driver'' function that
initializes data structures, defines the initial vertex for the
search, \etc.; this driver is not shown due to space limitations.

The depth-first search in {\small\verb+DFS_span+} proceeds as follows.  
If there are no fringe pairs to consider, we are done, and the (empty) 
fringe and spanning tree are returned to the caller.  If the
\emph{target} vertex is in the \emph{spanning_tree}, the search is
successful, and we return the fringe and spanning tree to the caller.
If there are fringe vertices to explore, we remove a \emph{(v, vpred)} 
pair from the front of the \emph{fringe} list (using the \emph{rst} 
rest-of-list function), then check to see if the vertex \emph{v} has 
already been encountered.  If so, we are done with this fringe pair, 
and so we  call {\small\verb+DFS_span+} recursively to access
the next fringe pair in the list (assuming there are any other pairs
left).  If \emph{v} has not been encountered before, we add the
\emph{(v, vpred)} key/value pair to the \emph{spanning_tree}, add
children nodes of \emph{v} to the \emph{fringe} list, utilizing the
\emph{explore} function (not shown), then call 
{\small\verb+DFS_span+} to continue the search.

This algorithm statement is compact, and (in our-not-so-objective 
opinion) quite elegant.  The dual use of the binary search tree for both 
marking visited vertices and recording the spanning tree as the
algorithm proceeds is especially satisfying.  Note also that a
breadth-first search implementation can be readily
obtained from the above by changing the \emph{explore} function
slightly, adding new fringe vertex pairs to the tail of the
\emph{fringe} list, rather than to the head of the list.  Note also
that a complete spanning tree from a given vertex can be obtained by
providing a ``bogus'' target vertex that doesn't actually exist in the
graph (we generally utilize the ``null'' vertex value of 0 for this
purpose).  Finally, the algorithm scales to millions of vertices, with
tens of edges per vertex, executes quite quickly, and is readily
compiled to hardware for even more speed.

\begin{figure}
\centering
\begin{boxedminipage}{1.0\textwidth}
{\footnotesize
\begin{verbatim}
function DFS_span (vtarget: in vertex, G: in graph,
                   spanning_tree: inout BST,
                   fringe: inout vertex_pair_list) {
  var v, vpred: vertex;
  in
    match fringe {
      -- Out of vertex pairs to process
      'Empty => skip;
      'Node n => {
        -- Found target vertex
        if exists(vtarget, spanning_tree) then
          skip;
        else {
          (v, vpred) := n.elt;
          rest(fringe);
          -- if v already found, on to the next fringe element
          if exists(v, spanning_tree) then
            DFS_span(vtarget, G, spanning_tree, vertex_pair_list);
          else {
            mark(v, vpred, spanning_tree);
            explore(MAX_EDGES, v, G, spanning_tree, vertex_pair_list);
            DFS_span(vtarget, G, spanning_tree, vertex_pair_list);
          }}}}}
\end{verbatim}
}
\end{boxedminipage}
\caption{Depth-First Search in DASL.}
\label{fig:dfs}
\end{figure}



\section{Use of ACL2 in DASL datatype and graphtype Design}
\label{RASLDASL}

The development of the DASL language, toolchain, and runtime is a
complex undertaking, and requires a solid foundation in the form of 
the basic \verb+datatype+ and \verb+graphtype+ design and 
implementation.  Early on, we identified the need for a rapid, yet
formal, prototyping environment that would allow us to experiment with 
design alternatives, evaluate these alternatives at scale, and provide
initial proofs of correctness for data structure implementations 
before committing to a final design.  Otherwise, we risked proceeding with 
flawed representations, requiring significant reworking of the DASL
toolchain to repair.  In short, we needed a 
``Semantic Laboratory'' for DASL development.  ACL2 filled the bill admirably: 

\begin{itemize}
\item{ACL2 is the most capable system we know of for the
    creation, proof, and execution of formal 
    specifications.  Using ACL2, we have been able to easily scale our data 
    structure prototype implementations to millions of vertices and 
    edges.  We have been able to quickly execute algorithms to exercise 
    these large data structures on concrete test input values, often
    generated at random, and validate the results of said
    algorithms operating on our prototype datatypes and graphtypes.}
\item{ACL2 provides \emph{single-threaded objects}, or \emph{stobjs},
    that provide functional data structure definitions with
    destructive ``under-the-hood'' implementations (subject to 
    basic syntactic restrictions that guarantee that no
    ``old'' versions of mutated data structures exist in a given
    function).  Thus, large data structures are
    not constantly being copied, as they would be for most pure
    functional implementations, eliminating garbage 
    generation/collection times for stobj data
    structures.  The contents of a stobj data structure also
    persist in the ACL2 world between events, providing convenient
    ``near-global'' variables that can be examined between
    function invocations, thus greatly aiding the debugging process.}
\item{ACL2 provides sophisticated proof libraries (books) for reasoning about
aggregate data structures of arbitrary size, as well as fixed-size
integers of various widths.}
\item{Tail recursion in ACL2 combines recursive functional style with 
    efficient compilation to loops.}
\item{ACL2 guards promote a type-like discipline with the added
    rigor of formal proof.  In particular, when we began 
    writing data structure mutators, it became obvious that, 
    in order to call said mutators from other guard-enabled functions, we
    would first need to prove that the mutators preserved the basic data
    structure ``footprint'' predicate provided by the \verb+defstobj+ event.}
\item{ACL2's simple packaging facility provides separate
    namespaces for datatypes/graphtypes.}
\item{All functions admitted to ACL2 must first be proven to
    terminate.  This encourages the ACL2 developer to explicitly consider termination
    issues when writing functions.}
\item{ACL2 is a mostly-automated theorem prover, and is quite 
    adept at automated inductive proofs.}
\end{itemize}
 
Thus, we implemented a Rudimentary ACL2 Semantic Laboratory for DASL,
which we cheekily refer to as RASL DASL.

Of course, not all of these ACL2 features work together 
without issue.  For example, it is more difficult to
perform proofs about tail-recursive functions than their
non-tail-recursive counterparts.  Guards are not part of the ACL2
logic; if one wants the benefit of guard predicates in a function,
one must restate those predicates in the function body.
Not all ACL2 books work well in concert.  Finally, stobjs are more 
difficult to reason about than, say, simple Lisp lists.  Indeed, 
when we first starting using tail recursion and stobjs 
to implement ``classic'' data structures, we were 
largely frustrated in our efforts to perform proofs about 
compositions of functions operating on stobj-based data
structures \cite{Hardin2009a}.  As often happens with ACL2, 
however, improvements in ACL2 over time, coupled with ACL2's 
uncanny ability to instruct the user to produce the kinds of forms 
that ACL2 ``likes'', have overcome many of these hurdles.  
We are now able to routinely obtain correctness proofs for 
compositions of tail-recursive functions operating on our 
prototype DASL ``in-place'' data structures using
stobjs, albeit with some user/prover interaction.

\subsection{Example datatype: Binary Search Tree}

As an example of the ACL2-based datatype prototyping effort, consider
the development of a basic Binary Search Tree (BST) datatype.  The
ACL2 stobj definition for this type is given in
Figure~\ref{fig:bst}.  (Note that graphtypes are defined similarly,
as the underlying data structure representation is explicitly designed to
represent graphs.)  The declaration of the \verb+Obj+ stobj is
followed by a number of basic functions (adding a vertex, deleting a
vertex, \etc. --- not shown due to space constraints), as well as 
\verb+defthm+ forms that provide basic lemmas about the components 
of the datatype (\eg, that the key array is an \verb+integer-listp+,
or that the result of updating the val array continues to satisfy the 
\verb+Objp+ predicate that is synthesized by the \verb+defstobj+ form 
depicted in Figure~\ref{fig:bst}).

\begin{figure}
\centering
\begin{boxedminipage}{1.0\textwidth}
{\footnotesize
\begin{verbatim}
(in-package "BST")

(defconst *MAX_VTX* 65535)
(defconst *MAX_VTX1* (1+ *MAX_VTX*)) ;; 2**16
(defconst *MAX_EDGES_PER_VTX* 2)
(defconst *MAX_EDGE* (* *MAX_VTX* *MAX_EDGES_PER_VTX*))
(defconst *MAX_EDGE1* (1+ (* *MAX_VTX* *MAX_EDGES_PER_VTX*)))
(defconst *MAX_EDGE_MINUS* (1+ (- *MAX_EDGE* *MAX_EDGES_PER_VTX*)))

(defstobj Obj
;; padding -- keeps ACL2 from turning (nth *VTXHD* Obj) into (car Obj)
  (pad :type t :initially 0)
  (vtxHd :type (integer 0 65535) :initially 0)
  (vtxTl :type (integer 0 65535) :initially 0)
  (vtxCount :type (integer 0 65535) :initially 0)
;; (V) This contains a pointer to the edge list for each vertex
  (vtxArr :type (array (integer 0 131069) (*MAX_VTX1*)) :initially 0)
;; (K) Keys for each vertex
  (keyArr :type (array (integer 0 *) (*MAX_VTX1*)) :initially 0)
;; (D) Data Value array
  (valArr :type (array (integer 0 *) (*MAX_VTX1*)) :initially 0)
;; (E) This contains pointers to the vertices that each edge is attached to
  (edgeArr :type (array (integer 0 65535) (*MAX_EDGE1*)) :initially 0)
 :inline t)
\end{verbatim}
}
\end{boxedminipage}
\caption{Binary Search Tree stobj declaration.}
\label{fig:bst}
\end{figure}

In DASL datatype/graphtype prototyping, we generally place the basic
structure definitions, \etc. in one file, and the higher-level
functions and theorems (\eg, functions that insert a new key/value 
pair, delete a key/value pair, check to see if a key is in the BST,
return the value for a given key, and so on) in a second file.  That way,
similar datatypes that share a common structure can reuse the file of basic
definitions, with slight modifications.  As an example of the sort of
higher-level functions we typically define, consider the
\verb+getVal+ function of Figure~\ref{fig:bstfn}, along with 
its shorthand macro form \verb+getV+.  \verb+getVal+ follows the
classic textbook definition, with a bit of the underlying ACL2 data
representation showing through.  Note that termination for this
function is not assured in general; thus, we add a \verb+count+ 
parameter that is decremented on each recursive call of
\verb+getVal+.  Fortunately, we have a convenient value at hand to
serve as a reasonable initial value for \verb+count+, namely 
the number of vertices in the BST.

\begin{figure}
\centering
\begin{boxedminipage}{1.0\textwidth}
{\footnotesize
\begin{verbatim}
(defun getVal (count key vtx Obj)
  (declare (xargs :stobjs Obj 
                  :guard (and (natp count) (natp key) (natp vtx))))
  (cond
   ((not (mbt (Objp Obj))) 0)    ;; Only positive values stored. 0 = 'null'
   ((not (mbt (natp count))) 0)
   ((not (mbt (natp key))) 0)
   ((not (mbt (natp vtx))) 0)
   ((zp count) 0)
   ((zp key) 0)
   ((zp vtx) 0)
   ((> vtx *MAX_VTX*) 0)
   ((mbe :logic (zp (vtxCount Obj))
         :exec (int= (vtxCount Obj) 0)) 0) ;; no vertices
   ((zp (keyArri vtx Obj)) 0)
   ((< key (keyArri vtx Obj))
    (getVal (1- count) key (edgeArri (left (vtxArri vtx Obj)) Obj) Obj))
   ((> key (keyArri vtx Obj))
    (getVal (1- count) key (edgeArri (right (vtxArri vtx Obj)) Obj) Obj))
   ;; (= key (keyArri vtx Obj))
   (t (valArri vtx Obj))))

(defmacro getV (key Obj)
  `(getVal (vtxCount ,Obj) ,key (vtxHd ,Obj) ,Obj))
\end{verbatim}
}
\end{boxedminipage}
\caption{Example DASL-derived Binary Search Tree function prototyped in ACL2.}
\label{fig:bstfn}
\end{figure}

\section{Results}

As an example of the sorts of algorithms we are able to successfully
prototype using the RASL DASL environment, consider the
hand-translated ACL2 version of the Depth-First Search algorithm of 
Figure~\ref{fig:dfs}, depicted in Figure~\ref{fig:acl2-dfs}.  The ACL2
version utilizes a graphtype  (\verb+gObj+, taking on the role of the 
DASL \verb+G+ parameter) and two datatypes: the Binary Search Tree
previously discussed (\verb+BST::Obj+, corresponding to the DASL 
\verb+spanning_tree+ parameter), and a doubly-linked list 
(\verb+DLST::Obj+, acting as the DASL \verb+fringe+ parameter).  
We employ a doubly-linked list, which provides access/update from 
either end of the list, for the \verb+fringe+ parameter because it was 
convenient to also use the same datatype for the implementation of the 
Breadth-FIrst Search algorithm (in fact, \verb+dfs_span+ and 
\verb+bfs_span+ are functionally identical; the difference in behavior 
between BFS and DFS is solely due to whether the \verb+explore+ 
function adds the vertex pairs it finds to the front (DFS) or back
(BFS) of the list.

\begin{figure}
\centering
\begin{boxedminipage}{1.0\textwidth}
{\footnotesize
\begin{verbatim}
(defun dfs_span (count vtarget gObj BST::Obj DLPR::Obj)
  (declare (xargs :stobjs (gObj BST::Obj DLPR::Obj)
                  :guard (and (natp count) (natp vtarget))))
  (cond
   ((not (and (mbt (natp count)) 
              (mbt (natp vtarget))
              (mbt (gObjp gObj)) 
              (mbt (BST::Objp BST::Obj))
              (mbt (DLPR::Objp DLPR::Obj)))) (mv BST::Obj DLPR::Obj))
   ((zp count) (mv BST::Obj DLPR::Obj))
   ((> vtarget *MAX_VTX*) (mv BST::Obj DLPR::Obj))
   ;; Found target vertex
   ((BST::existp vtarget BST::Obj) (mv BST::Obj DLPR::Obj))
   ((zp (DLPR::ln DLPR::Obj)) (mv BST::Obj DLPR::Obj))
   ;; Grab (v, vpred) pair from head of list
   (t (mv-let (v vpred) (DLPR::nthelem 0 DLPR::Obj)
        (cond
         ((not (posp v)) (mv BST::Obj DLPR::Obj))
         ((> v *MAX_VTX*) (mv BST::Obj DLPR::Obj))
         ((not (posp vpred)) (mv BST::Obj DLPR::Obj))
         ((> vpred *MAX_VTX*) (mv BST::Obj DLPR::Obj))
         ((BST::existp v BST::Obj)
          (seq2 BST::Obj DLPR::Obj
                (BST::nop BST::Obj)
                (DLPR::rst DLPR::Obj)
                (dfs_span (1- count) vtarget gObj BST::Obj DLPR::Obj)))
         (t (seq2 BST::Obj DLPR::Obj
                  (mark v vpred BST::Obj)
                  (seq DLPR::Obj
                       (DLPR::rst DLPR::Obj)
                       (explore *MAX_EDGES_PER_VTX* v gObj BST::Obj DLPR::Obj))
                  (dfs_span (1- count) vtarget gObj BST::Obj DLPR::Obj))))))))
\end{verbatim}
}
\end{boxedminipage}
\caption{Depth-First Search prototyped in ACL2.}
\label{fig:acl2-dfs}
\end{figure}

As in the BST \verb+getVal+ function, the termination of the
\verb+dfs_span+ function is not explicitly assured, so we again add a
\verb+count+ parameter that is decremented with each recursive call.
The number of vertices in the graph multiplied by the maximum number 
of edges per vertex serves as a reasonable initial value for \verb+count+.  
Another unique feature of the ACL2 code requiring some explanation is 
the \verb+seq2+ macro.  We often use J Moore's \verb+seq+ macro when 
working with stobjs, as it eliminates much of the \verb+let+ binding
``clutter'', allowing one to simply write one stobj-manipulating
expression after another within the scope of the \verb+seq+.  With 
two stobjs to update and return, the \verb+mv-let+ binding and 
\verb+mv+ return ``clutter'' can get much worse.  Thus, we developed a 
quick-and-dirty macro, \verb+seq2+ to do the analogous decluttering 
job for two stobjs.  \verb+seq2+ takes as parameters the two stobjs
that are to be updated, the updating s-expression for the first stobj,
the updating s-expression for the second stobj, and the s-expression 
that one wishes to return using \verb+mv+.  This works well enough in 
practice, although sometimes only one of the two stobjs one wishes to 
return using \verb+mv+ are actually modified within the scope of the 
\verb+seq2+.  In this case, one must provide some ``no-op'' 
s-expression for the unmodified stobj.

Other than these few changes, and a slight rearrangement to
concentrate the stobj updates into one contiguous section of the code, 
the ACL2 version tracks the DASL code fairly well, and the elegance of
the algorithm is still apparent.

To date, we have written a number of autonomy-relevant algorithms in 
DASL after prototyping in RASL DASL, including tree search, graph
search, Dijkstra's all-pairs shortest path algorithm, and
unify/substitute; as well as a number of supporting  
data structures, such as priority queues, stacks, singly- and
doubly-linked lists, queues, deques, \etc.  We have used our 
verification environment to state and prove properties of many of 
these algorithms, starting with basic well-formedness, and 
proceeding to full functional correctness.  Finally, we have 
generated code for many of these examples using the DASL 
compiler, and have validated the code generation via testing.  
As expected, the array-based form leads to efficient execution, 
scales well to millions of vertices with tens of edges per vertex, 
and enables hardware-based and GPU-based execution.

\section{Related Work and Future Work}

We first acknowledge the influence that ACL2
single-threaded objects \cite{stobj} have had on the overall 
DASL philosophical approach of providing functional specifications 
with efficient, ``under-the-hood'' implementations.  We 
have learned much from the ACL2 developers, particularly 
the importance of efficient, executable formal 
specifications.

The Boogie \cite{Boogie-IVL} and WhyML/Why3 \cite{why3} systems are
paradigmatic IVLs, supporting highly developed verification-enhanced
languages. The main point of departure with our work is that Boogie
and WhyML are programming languages, while the IVL for DASL is
higher order logic. Verification-enhanced versions of C are supported
by Boogie and Why3, and also by Appel's Hoare Logic for CompCert C
\cite{appel:prog-logic}, which is derived from the operational
semantics of CompCert in Coq.  The AutoCorres tool \cite {auto-corres} 
arose out of the \konst{seL4.verified} effort; it translates ASTs from
a parser for the C dialect used in seL4 to Schirmer's SIMPL theory in 
Isabelle/HOL and helps automate much of the Separation Logic used.  

Chlipala's Bedrock system \cite{bedrock} also utilizes higher order 
logic as an IVL; in particular, he uses Coq as the substrate on which 
to build intermediate and lower-level languages and prove the 
correctness of transformations.  Finally, the work of O'Leary and 
Russinoff on the formalization of C subsets for hardware design in 
ACL2 \cite{MASC} encouraged us to consider a formalized System 
language.

DASL is a direct descendent of the Guardol Domain-Specific Language
for Cross-Domain Systems \cite{guardol:tacas}.  Our experience with 
Guardol convinced us of the efficacy of taking a Domain-Aware approach 
to language development for various high-assurance domains.  Not 
surprisingly, DASL and Guardol share a number of syntactic and 
semantic features.  DASL also shares a great deal of the 
verification-oriented tool infrastructure pioneered on the Guardol 
effort, such as the logic-based IVL, the RADA backend tree solver, 
as well a VHDL code generator.  The latter capability allowed us to 
synthesize a formally proven high-level regular expression 
pattern matcher in inexpensive FPGA hardware that performed at 
Gigabit Ethernet line speeds \cite{GuardolVHDL}; these results
prompted us to continue to develop VHDL code generation for DASL. 
DASL distinguishes itself from Guardol mainly in the 
\verb+datatype+ and \verb+graphtype+ declarations, the \verb+sized+ 
declaration, and the attendant syntactic restrictions that allow for 
compilation to an efficient in-place data structure representation.

In future work, we will continue to develop the DASL toolchain,
focusing on code generation for CakeML, but also improving code
generation for Ada, Java, VHDL, and other target languages.  We will also
continue to refine the language syntax, focusing on ways to 
conveniently add extensions such as grammar specifications 
and rulebases.  Regular expressions are introduced currently by way of 
a simple \verb+regex_match+ intrinsic function, which accepts a 
regular expression as a string parameter, but this method is unlikely 
to scale well to grammar rules, \etc.  Finally, we will continue to expand
on the suite of applications implemented in DASL, especially in the
areas of autonomy algorithms, as well as data filtering/transformation 
for cyber-resilient systems.

\section{Conclusion}

We have developed a verifying compilation technique for a domain-aware
programming language for autonomy and other high-assurance
applications that combines efficient
implementations of high-level data structures used in autonomous
systems with the high assurance needed for accreditation.  Our system
supports a ``natural'' functional proof style, yet applies to more
efficient data structure implementations.  Our toolchain features code
generation to mainstream programming languages, as well as GPU-based
and hardware-based realizations.  We base the Intermediate
Verification Language for our toolchain upon higher-order logic.  By
giving program execution a formal semantics, claims about program
behavior can be mathematically proven, and source-to-source
transformations of the intermediate form can be proven as well. 
Thus, proofs about high-level programs over high-level data 
structures can be carried out while automatically ensuring a formal, 
proved connection to the low-level efficient implementation.  
We have also demonstrated that when the IVL is
higher order logic, verified code generation is possible via the
facilities of the CakeML verified compiler.

We utilized ACL2 to develop our efficient yet verifiable data structure
design.  ACL2 is  particularly well-suited for this task, with its 
sophisticated libraries for reasoning about aggregate data structures 
of arbitrary size, efficient execution of formal specifications, 
and its support for single-threaded objects, not to mention its
strength as an automated inductive prover.  We described our 
high-assurance data structure design approach in ACL2, 
presented ACL2 examples of common algebraic data types implemented 
using this design approach, discussed proofs of correctness for those 
data types carried out in ACL2, as well as sample ACL2 implementations 
of relevant algorithms that utilize these efficient, high-assurance data
structures.  In summary, this ACL2-based development activity has 
produced a performant, as well as highly-assured, data structure
design for critical applications, and we continue to use this
ACL2-based design environment to prototype new applications for 
modern high-assurance systems, such as lexers and parsers for 
data interchange formats such as JSON, inference engines, and the like.

\section{Acknowledgments}

We thank the anonymous referees for their helpful comments.
This work was sponsored in part by the Defense Advanced 
Research Projects Agency (DARPA).

\bibliographystyle{eptcs} 
\bibliography{biblio}

\end{document}